# NUMERICAL METHODS FOR PULMONARY IMAGE REGISTRATION

Roberto Cavoretto, Alessandra De Rossi, Roberta Freda, Hanli Qiao, Ezio Venturino

Department of Mathematics "G. Peano", University of Turin, via Carlo Alberto 10, 10123 Torino, Italy

Email: *roberto.cavoretto@unito.it*

**Abstract.**

**Introduction:** Due to complexity and invisibility of human organs, diagnosticians need to analyze medical images to determine where the lesion region is, and which kind of disease is, in order to make precise diagnoses. For satisfying clinical purposes through analyzing medical images, registration plays an essential role. For instance, in Image-Guided Interventions (IGI) and computer-aided surgeries, patient anatomy is registered to preoperative images to guide surgeons complete procedures. Medical image registration is also very useful in surgical planning, monitoring disease progression and for atlas construction. Due to the significance, the theories, methods, and implementation method of image registration constitute fundamental knowledge in educational training for medical specialists. In this chapter, we focus on image registration of a specific human organ, i.e. the lung, which is prone to be lesioned. For pulmonary image registration, the improvement of the accuracy and how to obtain it in order to achieve clinical purposes represents an important problem which should seriously be addressed. In this chapter, we provide a survey which focuses on the role of image registration in educational training together with the state-of-the-art of pulmonary image registration.

**Methods:** In the first part, we describe clinical applications of image registration introducing artificial organs in Simulation-based Education. In the second part, we summarize the common methods used in pulmonary image registration and analyze popular papers to obtain a survey of pulmonary image registration.

**Conclusions**: The study, design and employment of artificial lungs in training, together with pulmonary image registration and clinical applications of image registration, have reached a certain level of maturity. However, since the field is rapidly developing, the reader can find in this chapter the state-of-the-art, which constitutes a useful basis to understand the future progresses in this field.

# 1    Introduction

**I**mage registration is the process that aligns a pair of images to the common spatial coordinates. These images might be obtained by various modalities, viewpoints, or at different times. There are many fields in applied sciences and engineering that use image registration. Especially in medical imaging, it is helpful for clinicians to analyze, extract and fuse information from various images. In this chapter, we focus on medical image registration on a specific human organ, the lung.

Lung is primarily used to respiration, hence it is one of the most important organs in humans. But it is prone to be lesioned, when people have lung diseases, such as cancer, pulmonary shadow or embolism, respiratory failure, *et al*. For diagnosing the local diseased part, physicians need to perform an analysis of pulmonary images taken by specific modalities. If diagnosticians skip registration, they need to analyze and fuse information from the images in their minds, basing on their knowledge. This means that physicians must possess very rich and broad experiences. The invisible human anatomy causes a hard challenge for them to obtain precise diagnosis without medical image analysis.

Another important content in medical science is educational training. As we all know, for medical specialists, physicians, diagnosticians and health care workers, educational and professional training is essential. Beyond learning the theories, clinical training plays the crucial role to incorporate knowledge into practice. Therefore, suitable learning methods should be established. With the increasing number of students and in order to ensure patient safety, the traditional methods such as providing clinical skills and experiences for students in hospitals no longer meet the needs of the curriculum [1]. To satisfy the practical purpose, Simulation-Based Education (SBE) has become a prominent method for clinical training.

This chapter does not present novel strategies of clinical applications and numerical methods for pulmonary image registration, but it is a review of the analysis, a summary and a comparison of various algorithms. We divide this chapter into two parts: in the first part we discuss the clinical applications of synthetic organs and image registration, while the second part mainly focuses on numerical methods for pulmonary image registration.

## 2    Artificial Lung and Image Registration in Clinical Applications

SBE is one of the most effective ways for improving skills and practical knowledge of medical students. It is also helpful for medical staffs in hospitals to consolidate certain clinical scenarios. A special case in SBE for surgeons is surgery training. In order to avoid injuries of animals, the use of artificial organ simulator has occupied an increasing important role to satisfy the learning purposes.

### 2.1    Artificial Organs

Due to the challenges in finding donors and matching organs for patients in need of a transplant, a quick and simple process to get new organs should be available. Using artificial organs to accomplish transplant surgeries for surgeons is an efficient technology, since they act in the same way as natural organs. These organs are composed of large synthetic tissues that are very similar to the make-up of real organs and function in the same way [2]. Beyond the utility in transplant surgeries, there are other reasons to justify the use of artificial organs. For disabled patients, artificial organs, such as artificial limbs, can improve their ability for self care. Fig. 1 shows an artificial heart valve [2]. It depicts another goal of scientists in creating artificial organs. Focusing on the lung, Federspiel and Svitek in [3-4] introduced the artificial lung from various viewpoints: principles, applications and future directions. The purpose of creating the artificial lung is to replace or supplement the respiratory function of diseased lungs. There are three next generation artificial lungs: 1) paracorporeal; 2) intravascular and 3) intrathoracic artificial lungs. For these various types of artificial lungs, the corresponding treatments are different. Precisely, paracorporeal and intrathoracic artificial lungs are suitable for bridge-to-transplant respiratory support, whereas the intravascular ones are applied in case of acute respiratory failures.

### 2.2    Artificial Organs Models in SBE

For eliminating the use of human tissues, animal testing, and cadavers for surgical training and medical education, designing human organ models is a useful strategy for trainees [5]. Synthetic organ is an artificial organ manufactured by humans. One of the advantages of the synthetic human organ models is that they can replace the use of human tissues, cadavers and animals testing in surgery training for surgeons and medical students. In this way, the trainees can replicate procedures many times to improve and consolidate their skills. A special synthetic human organ modeled by using silicone gel is proposed by the *University of Minnesota* [5]. The tool has been created to produce organ-specific computer-generated three dimensional models. Firstly, plastic shapes are produced by 3D printing technique and then the silicone gel is used to obtain final model. The latter is tested against a human tissue database to determine its accuracy, thereby enabling to guarantee its efficiency in educational training. For clinical surgical learning, designing more accurate and realistic human organ models in order to exploit their advantages represents the future direction to go.

Scientists have been struggling with designing a specific artificial lung due to the large number of complex tissues. For instance, recently, scientists belonging to a research group of the *Technion-Israel Institute of Technology* have developed a life-size artificial lung to better understand how aerosols impact human health, [6]. Owing to the small size of the aerosols and of the complexity of lung tissues, the movement of aerosols in the respiratory system is very difficult to monitor. The diagnostic tool in this paper is the first strategy that enables quantitative monitoring of the dynamics of aerosols at such small scales.

For satisfying educational needs, the existing lung simulators usually cost a lot and are complicated to be incorporated in a simulation room. For this reason, a pediatric and neonatal lung simulator has been created by Díaz Boladeras, et al. [7]. The simulator proposes an innovatory ventilation for fulfilling educational purposes through a compact, simple and low-cost product.

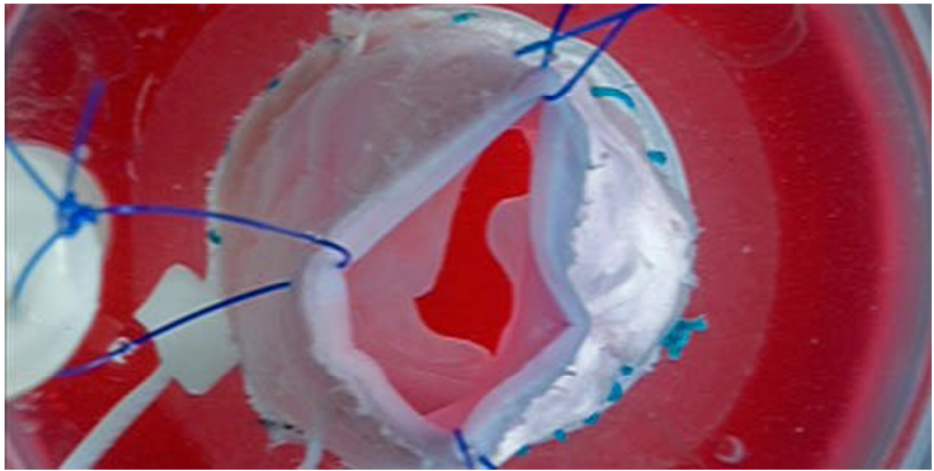

Fig. 1. Artificial heart valve

### 2.3 Image Registration in Clinical Applications

Image registration can be used in numerous clinical applications, such as correct delineation of lesions. In particular, automated image registration can help diagnoses to make the final manual report for patients. This is the reason for which the hospital staffs need to be trained to learn the techniques and the related knowledge. Image registration methods, publications on the topic and clinical applications are recalled and reviewed in [8]. In medical science, the most modern use of image registration is the Image-Guided Surgery (IGS). The latter is a general term to describe any surgical procedure, where the surgeon employs tracked surgical instruments in conjunction with preoperative or intraoperative images in order to indirectly guide the procedure [9]. During surgeries, safe and less invasive procedures can be achieved through IGS. In this field, the performance of image registration is essential since the more precise registered results are, the better surgical outcomes are. For example, [10] presents a nonrigid registration method in image-guided MR-TRUS prostate intervention, which is a good example of the application of image registration in IGI.

Three medical application fields of IGI and corresponding instances are presented in [11]. They are interventional radiology, surgery and radiation therapy, respectively. The main tools in interventional radiology are needles and catheters. Traditionally, the guidance for interventional radiologists has relied upon intraoperative images, such as X-ray fluoroscopy. But the images are 2D projections, and this is a disadvantage for radiologists since they need guidance in the 3D working space. This paper proposes a solution to enhance the intraoperative images with preoperative images under the fact that preoperative MR or CT images contain high-resolution 3D information. By registering the preoperative images to the intraoperative images, the 3D information can be stored to improve the guidance. In interventional surgery, there are two routes for surgeons in order to make procedures less invasive: endoscopic surgery and navigation systems. A first registration step is required to align preoperative images and apply magnetic tracking to monitor the corresponding positions of instruments during the surgery.

The workflow of IGI is depicted as follows [12]:

- Acquisition of the preoperative images
- The positions of instruments are tracked using a localizer
- During the procedure, the positions of instruments are displayed on the image relative to the patient anatomy
- Surgeons use the virtual display to accomplish the procedure
- A confirming image is obtained upon procedure completion

Image guidance enables Minimally Invasive Procedure (MIP). Nowadays, MIP has become the main surgery trend. The advantages of MIP reside in the less painful, more cosmetic incisions and in the shorter hospital stay of the patients. MIP can also decrease the morbidity and mortality. For a better knowledge about MIP and to rehearse the operative skills, trainees and medical students should follow the specific projects, such as the S*urgical Simulation*, *Planning and Image Guided Surgery* from Center for Biomedical Technology [13]. In such projects, image registration represents a fundamental knowledge and therefore needs to be taught.

Image registration is not only the main technology in IGI, but it can also be applied to patient-specific virtual physiological human (VPH) models. The survey [14] highlights the existing techniques, one of which is used for patient-specific image registration in order to extract different classes of tissues. In [15], Minimally Invasive Lung Transplantation (MILT) and Clamshell Incision (CS) are comparatively analyzed through various data. The authors conclude that despite MILT takes longer ischaemic times and needs more complex operations, compared with the traditional routine approach of CS, it still performs early preoperative and mid-term clinical benefits.

## 3     A Survey of Pulmonary Image Registration Methods

In modern medicine, owing to the complexity and invisibility of human organs, image analysis has become a very prominent topic. Particularly, in the field of pulmonary science, Computed Tomography (CT) image analysis plays a key role in clinical applications to diagnosis, treatment and monitoring.

During image analysis, for many of these applications it is necessary to determine the spatial correspondence between two pairs of images. The process to establish the correspondence is called image registration, which is an efficient strategy to analyze and visualize human anatomy in order to satisfy clinical purposes. Registration allows an alignment of images and the combination of information coming from various sources, or from multiple scans with different modalities. It is also useful to correct the deformation of a patient's anatomy over time (e.g. correct the motions of preoperative, intraoperative and postoperative images), in order to estimate the physiological motions of organs and monitor the disease progression. For pulmonary image analysis, registration is able to extract quantitative values of the type and growth of pulmonary nodules for early diagnose malignant neoplasms. Furthermore, in surgical planning and IGI, pulmonary CT image registration is used to estimate breathing motion, monitor success of treatments, assess ventilation, quantify gas exchange. Moreover, it is also useful for diagnosing fibrosis or Chronic Obstructive Pulmonary Diseases (COPD), such as emphysema.

In general, image registration can be described as the following process: given two $d$-dimensional images called *source image* and *target image* (usually $d$ equals to 2, 3 or 4), the goal is to find a suitable transformation between two images such that the transformed target image is similar to the source image. According to various criteria, image registration can be classified as different types, such as intensity- and feature-based algorithms, or depending on the nature of the transformation, the registration algorithms are rigid, affine, projective, or elastic. Registrations are implemented in various modalities and are sorted into single- or multi-modality approaches. For an overview, see the references [16-20, 37, 38]. Often, to find the transformation, it is common to minimize a certain energy functional (sometimes also called the *cost function*) composed by two building blocks. One of them is a distance measure between the source and the transformed target images, and represents the external forces, which are computed from the image data to obtain the desired registration result. The second one is a regularization condition that computes internal forces, and it is added to render the transformation physiologically plausible and to avoid discontinuities like gaps or foldings.

Hence, the task in a registration problem is to find the best possible transformation that defines a point-to-point correspondence between the two images and minimizes the cost function. In this section, we present common algorithms used in pulmonary image registration and a survey of the corresponding methodologies.

### 3.1 Common Algorithms for Pulmonary Image Registration

Focusing on pulmonary image registration, the most common approaches are intensity-, feature-, segmentation-based, mass-preserving and diffeomorphismic algorithms. We will introduce these methods in the subsequent paragraphs.

#### 3.1.1 Intensity-based Registration Techniques

Intensity-based registration techniques are methodologies based on voxel intensity information over complete images. They depend on mathematical or statistical criteria to match intensity patterns between the deforming scan and the source image. Generally, intensity-based registration algorithms produce more accurate results to improve the overall accuracy of the alignment. Intensity-based registration is a very popular method, for instances see [21-24] for the applications to pulmonary images.

As mentioned, two compositions of transformations are the distance measure and the regularizer. Here we report the most common regularizers in case of a dense displacement field.

- *Diffusion.* It is the physical process that equilibrates concentration differences without creating or destroying mass. Even if it is not strictly related to the physical motivation, diffusion filtering is commonly used as a technique for smoothing images. It also allows a simple and efficient implementation [25-27].
- *Elastic body.* In this approach, the registration process simulates a model in which one of the images, made from an elastic material, is deformed until it matches the other image. The optimal mapping is obtained by finding the equilibrium state of the elastic image, which corresponds to a local minimum of the total energy. The equilibrium state is obtained by solving a set of partial differential equations derived by the linear theory of elasticity. Often the image domain is modeled as an elastic body based on the Navier-Cauchy PDE that tries to balance the external forces with the internal stresses imposing smoothness constraints [25-28].
- *Viscous flow.* This approach is based on the Navier-Stokes equations and the image domain is modelled as a viscous fluid that behaves as a collection of particles conforming to Newtonian mechanics [28]. The Navier-Stokes PDEs describe the balance of forces acting in a given region of the fluid. In this model the transformations are obtained by numerically solving a set of nonlinear PDEs associated with the constrained optimization problem. This approach is often useful when images with large deformations have to be registered, for example in the case of intersubject registration where the anatomical variation over a population has to be taken into account. A drawback of this model is the computational inefficiency, so most developments try to improve its numerical performance [26], [28].
- *Curvature*. Fischer and Modersitzki in [27] introduced a new curvature-based registration technique in which the internal forces are designed to minimize the curvature of the displacement field, that is approximated by an integral. These internal forces provide smooth solutions and allow automatic affine linear alignment; consequently, a pre-registration step is not necessary.

Once the regularizer is chosen, to complete the construction of the function, also a distance measure has to be selected.

Usually image similarity is measured by regarding the intensity differences in each voxel. There are a lot of distance measures that can be used in image registration. The most common in lung image registration are the following ones:

- **SSD**. This distance measure is the Sum of Squared Differences between the intensity voxel of the target image and the intensity in the transformed one. This measure gives satisfactory results if the intensities of the two given images are comparable, with at most some Gaussian noise. Since this, it is used mainly in the case of mono-modal registration.
- **NCC**. Normalized Cross Correlation is a good similarity measure if there is a linear relationship between the intensity values in the source and template image [18]. Cross correlation is invariant to the linear intensity change. Because of this, it is used for multimodal registration but also to overcome intensity changes due to tissue compression in lung image registration.
- **Mutual Information**. It can be used for multi-modal registration and also for lung alignment. Viola and Wells in [29] present an algorithm that aligns images by maximizing the mutual information between source and target images. It maximizes the dependence between images minimizing the joint entropy of the intensity values of the images. The mutual information between target and transformed source images has three components. The first one is the entropy of the target image. The second one is the entropy of the part of the transformed source image. The third one, is the (negative) joint entropy of the two images. It contributes when the two images are functionally related. We observe that the entropies are defined in terms of integrals over the probability densities associated with the random variables, [29]. However, analyzing signals or images, often we do not have direct access to the densities. Because of this, also a differentiable estimate of the entropy of a random variable, calculated from samples, is given.
- **Normalized Gradient Field.** Often mutual information has several local minima and needs the probability densities that often are not accessible. To overcome these drawbacks, Haber and Modersitzki in [30] proposed to align images exploiting their gradients rather than their intensities. This approach is simpler, fast to implement and more suitable to optimization. It is based on the facts that two images are considered "similar" if intensity changes occur at the same locations; moreover image intensity changes can be detected using the image gradient. The NGF is not differentiable in areas where the image is constant and highly sensitive to small values of the gradient field. Because of this, in [30] a regularized normalized gradient field is also defined, in which a parameter that represents the minimal size of a local change in the image is introduced.

### 3.1.2 Mass Preserving Registration Algorithm

In numerous approaches, some specific properties of lung anatomy are modeled as an additional distance metric, for instance, mass-preserving. The lung appearance in CT depends significantly on the amount of the air inhaled. Firstly, because the lungs are larger in size at the inspiration level, and secondly because the lung tissue saturates additional air and appears darker in CT scans, which should not be confused with the emphysema progression and lung tissue destruction. Because of this, in [22] and [31] a method in which the mass of lungs is preserved during the breathing cycle is proposed.

The mass preserving model describes the change in density in lung CT scans related to the change in lung volume. In fact, at the base of this model there is the fact that the density of the lung tissue is inversely proportional to the local volume. Hence, changes in the density produce changes in the volume and vice versa. Furthermore, the absolute value of the Jacobian determinant of the transformation is related to the expansion (or contraction) of the lung volume (and therefore to the density of the lung tissue). The idea is to model the lung tissue density using the determinant of the Jacobian of the transformation and to modify the sum of the squared differences similarity function to allow mass preservation and to simulate how the lung tissue appears under the given deformations. The mass-preserving methodology is incorporated into a framework of multi-resolution image registration in which the images are firstly registered with a global affine transformation, and then a series of B-Spline transformations is utilized with decreasing grid size. The mass-preserving image registration performs better than other methods without mass preservation assumption in image pairs with a considerable difference in inspiration level.

### 3.1.3 Feature-based Registration Techniques

Feature-based registrations are the techniques that rely upon geometrical structures extracted from the original images. These features are low-dimensional, and require a representation of the image data in terms of distinctive geometrical structures. These features could be, for example, points, curves or surfaces, which correspond to landmark-, curve- and surface-based registration, respectively. For a finite number of features the idea is to determine a transformation that maps each feature of the target image into the corresponding feature of the source one [25].

Feature-based algorithms offer more robust registration when image intensity is changed, for example in the presence of pathologies, image artifacts or differences in scan protocol. On the other hand, major drawback of feature-based registration methods is the necessity to extract reliable features with more accurate estimate deformation fields of the extracting features. Based on the benefits of intensity- and feature-based methods, for improving the register accuracy and overcoming the drawbacks, some researchers have proposed hybrid methods combining intensity- and feature-based algorithms [22].

### 3.1.4 Landmark-based Registration

Landmarks can be anatomical or geometrical. There are two ways to detect the landmarks. First, by using salient and accurately located points of the anatomical morphology, usually identified interactively by the user. Secondly, geometrical landmarks are points at the locus of the optimum of some geometric property, generally localized in an automatic fashion [20].

In theory, landmark-based registration is versatile, which means that it can be applied to any image, independently of the object or subject. At first, landmark-based methods were mostly used to find rigid or affine transformations; nowadays, some scientists have employed landmark-based image registration for assessing elastic transformations. Anatomical landmarks are also often used in combination with entirely different registration bases; for instance in [24] an intensity- and landmark-based registration algorithm is proposed.

### 3.1.5 Curve and Surface-based Registration

Curve and surface-based registrations use distinctive anatomical lung structures such as vessels and lung surfaces as features for registration. These structures have to be segmented from the image before starting to register it.

In [22] a curve and surface-based registration method is presented. Lung surfaces and vessel tree centrelines are built according to the *current-based registration* and aligned using the metric on currents.

Geometrical shapes such as curves and surfaces are represented with a set of vectors. A Current is defined with a finite set of vectors attached to specified positions; the curve can then be defined with its tangent vector at each position. In the case of a discrete setting, a curve is considered as a set of piecewise linear segments, where each segment is represented by its center point, tangential direction, and segment length. In the same way, a surface is defined by the normal direction, the face center and the area.

### 3.1.6 Segmentation-based Registration Methods

Actually, segmentation is the operation of preregistration process. Segmentation-based registration can be rigid model or deformable model based. For the rigid model based case, the same structures from both images to be registered are anatomically extracted, and used as a unique input for the alignment procedure. In the deformable model based case, an extracted structure from one image is elastically deformed to fit the second image. Due to the simplicity of implementation and relative low computational cost for segmentation, the method has remained popular, and many follow-up papers aimed at rendering the segmentation step automatic in order to improve and optimize the performance. However, the drawback of segmentation-based methods is the accuracy of the registration, that is limited by the accuracy of the segmentation. An integrated segmentation and registration method is proposed in [26] for improving the register accuracy in pulmonary CT images.

### 3.1.7  Diffeomorphisms

Diffeomorphisms (i.e. differentiable mappings with differentiable inverses) are frequently used in medical image registration because they preserve the topology: disjoint sets remain disjoint and the smoothness of the anatomical features is preserved. A diffeomorphic transformation can be obtained integrating a time-dependent, smooth velocity field through an ordinary differential equation. It can be time- and memory-consuming, because of this, the restriction to stationary velocity fields is also examined, considering a diffeomorphism generated by an exponential mapping. For details about this topic see [26], [32].

## 3.2  State-of-the-art of Pulmonary Image Registration

In 3.1, we summarized the most popular techniques for pulmonary image registration. In this subsection, we present from various viewpoints, i.e. image data, purposes, strategies, and comments presented in some significant papers.

In [21], the topic of regional pulmonary function analysis is considered. For evaluating the reproducibility of regional pulmonary function, 4D CT and image registration are used. To establish reproducibility of various ventilation measurements in 4D CT image data, the author develops and validates a process by computing the determinant of the deformed field Jacobian matrix. The lung CT image registration problem is used to monitor emphysema progression in [22]. For solving this problem, a mass preserving model based registration method is proposed. A feature- and intensity-based image registration methods are also presented to improve the register accuracy. Also, the framework for monitoring emphysema progression is developed. The goal of [23] is instead to determine whether a standard and generic, but fully automatic, intensity-based image registration algorithm can achieve comparable results. The image data are CT images provided by EMPIRE10 [36] and the algorithm is implemented in *elastix*. For tracking tissue deformation during the respiratory cycle, in [24], the authors propose an intensity- and landmark-based image registration algorithm under consistency constraints, matching corresponding airway branchpoints, in order to perform image registration and warping of 3D pulmonary CT image data. The proposed method is demonstrated to be able to reduce the average landmark registration error, which potentially provides the accurate solution of the problem of tracking 3D internal lung structures during a respiratory cycle. Alexander Schmidt-Richberg, in [26], presents a new integrated registration and segmentation of lung and lung lobes in 3D and 4D CT images. The author mainly inspects how registration benefits from an explicit consideration of lung-specific morphological and physiological characteristics which are modeled by segmentation. The integrated method is used in the first part of this book to explicitly align the pulmonary lobes. The second part instead elaborates the physiological properties of the breathing motion at the lung boundaries by segmentation. Evaluations and results obtained from EMPIRE10 [36] show that the segmentation-based considerations greatly improve registration accuracy and plausibility. During the breathing cycle, in order to estimate the local change in lung tissue intensity, a mass-preserving registration technique is proposed in [31] on a large number of CT images. The method is an incorporation between a tissue appearance model based on the preservation of total lung mass and a deformable registration framework. The evaluated results show that the image registration method under mass preservation performs significantly better than the image registration approaches without such constrain, with a considerable difference in the inspiration level. Especially, the mass preserving assumption helps to obtain physically plausible deformations in regions without strong image gradients. In [32], the two diffeomorphic transformation models, greedy Symmetric Normalization and exponential mappings are applied to image data provided by EMPIRE10. The whole image registration pipeline is built on ANTS, an open-source toolbox. The results obtained from EMPIRE10 among 34 total algorithms show that the methods have state-of-the-art performance. Thesis [33] mainly evaluates deformable lung registration for pulmonary image analysis in MRI and CT scans. A comprehensive framework combining some novel mathematical formulations and computational solutions is presented. Moreover, a new metric, called *Textural Mutual Information* (TMI), is proposed. It is shown to be robust against certain intensity distortions. For overcoming such computational problems, the concept of *Modality Independent Neighbourhood Descripto*r (MIND) was introduced. For both multi-modal registration (MRI and CT) and single-modal motion estimation, a significant improvement in terms of accuracy of this methodology using MIND is compared to conventional similarity metrics. To remove the restrictions of the underlying registration model, a graph-based optimization procedure is proposed. For solving lung CT matching problems, the author of [34] introduces a new volumetric registration algorithm and also presents a novel registration scheme utilizing both surface and volumetric algorithms for overcoming mismatching in some complex registration problems. Finally, in [35] three measures, namely SAJ, SACJ, SAI are compared with each other to estimate regional lung tissue ventilation. The ventilation comes from tissue volume and vesselness-preserving image registration in CT images. The three registration-based ventilation measures are also compared to Xe-CT estimates of specific ventilation (Xe-CT sV) measurement.

## 3.3   EMPIRE10

The popular and public platform EMPIRE10 (Evaluation of Methods for Pulmonary Image REgistration 2010) challenge is able to evaluate the state-of-the-art in lung CT image registration. The EMPIRE10 challenge was launched in early 2010, followed in September by a workshop at the MICCAI 2010 conference. The main goal of EMPIRE10 is to fairly compare image registration algorithms on a same database of intra-patient thoracic CT pairs. EMPIRE10 provides a unique opportunity to make legitimate comparisons among various registration algorithms. A detailed discussion about EMPIRE10 challenge can be found in [36].

There are four separate ways to evaluate the submitted registration results: alignment of the lung boundaries, alignment of the major fissures, correspondence of annotated point pairs and analysis of singularities in the deformation field. In all these cases, after finishing the evaluation for determining their placements, each research team will receive a score: the lower it is, the better it is.

## 4     Conclusions

The study, design and employment of artificial lungs in training and Simulation-based Education, together with pulmonary image registration and clinical applications of image registration are here reviewed. They have reached a substantial level of maturity. The reader can find in this chapter the present state-of-the-art of the field, with a survey of the many methods, techniques and software employed in it. However, since the field is rapidly developing, this presentation can be retained also later on as a starting point useful to understand further progresses.

# References

1. Weller JM, Neste lD, Marshall SD, Brooks PM, Conn JJ. Simulation in clinical teaching and learning. *The Medical Journal of Australia* 2012; 196(9): 594-598.

2. Synthetic Organs: The future.

   http://synthetic-organs.yolasite.com/synthetic-organs.php

3. Federspiel WJ, Svitek RG. Lung, Artificial: Current Research and Future Directions. *Encyclopedia of Biomaterials and Biomedical Engineering* 2004; 922-931.

4. Federspiel WJ, Svitek RG. Lung, Artificial: Basic Principles and Current Applications. *Encyclopedia of Biomaterials and Biomedical Engineering* 2004; 910-921.

5. Special Order Synthetic Human Organ Models for Surgery Training.

   http://license.umn.edu/technologies/20110220-5_special-order-synthetic-human-organ-

   models-for-surgery-training

6. Fishler R, Hofemeier P, Etzion Y, Dobuwski Y, Sznitman J. Particle dynamics and deposition in true-scale pulmonary acinar models. *Scientific Reports* 2015; 5 14071, doi: 10.1038/srep14071.

7. Abdullah Z, Basiuras A, Duman A, Viscarri L. Pediatric and Neonatal Lung Simulator. European Project Semester. June 2014.

8. Imran MB, Meo SA, Yousuf M, Othman S, Shahid A. Medical Image Registration: Basic Science and Clinical Implications. *J Ayub Med Coll Abbottabad* 2010; 22(2): 199-204.

9. Mezger U, Jendrewski C, Bartels M. Navigation in surgery. *Langenbecks Arch Surg* 2013; 398: 501-514.

10. Onofrey J, Staib LH, Sarkar S, Venkataraman R, Papademetris X. Learning nonrigid deformations for constrained point-based registration for image-guided MR-TRUS prostate intervention. IEEE, 12th International Symposium on Biomedical Imaging (ISBI) 2015; 1592-1595, doi: 10.1109/ISBI.2015.7164184.

11. Sauer F. Image registration: enabling technology for image-guided surgery and therapy. IEEE, Engineering in Medicine and Biology 27th Annual Conference 2005; 7242-7245.

12. Cleary K, Peters TM. Image-guided interventions: technology review and clinical applications. *Annual Reviews of Biomedical Engineering* 2010; 12: 119-142.

13. Surgical Simulation, Planning and Image Guided Surger

   http://www.ctb.upm.es/?page_id=365

14. De Oliveira JEE, Giessler P, Deserno TM. Image registration methods for patient-specific virtual physiological human models. VCBM 15: Eurographics Workshop on Visual Computing for Biology and Medicine 2015; 31-40.

15. Marczin N, Popov A-F, Zych B, Romano R, Kiss R, Sabashnikov A et al. Outcomes of minimally invasive lung transplantation in a single centre: the routine approach for the future or do we still need clamshell incision? *Interact CardioVasc Thorac Surg* 2016; 22: 537-545.

16. Ardeshir Goshtasby A. 2-D and 3-D image registration for medical, remote sensing, and industrial applications. Bibliometrics, WILEY, 2005.

17. Oliveira FPM, Tavares JMRS. Medical Image Registration: A Review. *Computer Methods in Biomechanics and Biomedical Engineering* 2014; 17(2): 73-93.

18. Hill DLG, Batchelor PG, Holden M, Hawkes DJ. Medical Image Registration. *Physics in Medicine and Biology* 2001; 46: R1-R45.

19. Sotiras A, Christos D, Paragios N. Deformable Medical Image Registration: A Survey. Research Report 2012, RR-7919.

20. Maintz JBA, Viergever MA. An Overview of Medical Image Registration Methods. *Medical Image Analysis* 1998; 2(1): 1-36.

21. Du K.F. Regional pulmonary function analysis using image registration and 4D CT. PhD thesis, University of Iowa 2013.

22. Gorbunova V. Image registration of lung CT scans for monitoring disease progression. PhD thesis, Copenhagen University 2010.

23. Staring M, Klein S, Reiber JHC, et al. Pulmonary Image Registration with Elastix using a Standard Intensity-Based Algorithm. In: van Ginneken B, Murphy K, Heimann T, et al, eds. *Medical. Image Analysis for the Clinic: A Grand Challenge* 2010; 73-79.

24. Li BJ, Christensen GE, Hoffman EA, McLennan G, Reinhardt JM. Pulmonary CT image registration and warping for tracking tissue deformation during the respiratory cycle through 3D consistent image registration. *American Association of Physicists in Medicine* 2008; 35(12): 5575-5583, doi: 10.1118/1.3005633.

25. Modersitzki J. Numerical Methods for Image Registration, Oxford University Press, Oxford, 2004.

26. Schmidt-Richberg A. Registration Methods for Pulmonary Image Analysis, Integration of Morphological and Physiological Knowledge, Springer Fachmedien Wiesbaden 2014; doi: 10.1007/978-3-658-01662-3_1.

27. Fischer B, Modersitzki J. A unified approach to fast image registration and a new curvature based registration technique. *Linear Algebra and its Applications* 2004; 380: 107-124.

28. Holden M. A review of geometric transformations for nonrigid body registration. IEEE Transactions on Medical Imaging 2008; 27(1): 111-128.

29. Viola P, Wells III WM. Alignment by Maximization of Mutual Information. *International Journal of Computer Vision* 1997; 24: 137-154.

30. Haber E, Modersitzki J. Intensity gradient based registration and fusion of multi-modal images. In: Medical Image Computing and Computer-Assisted Intervention (MICCAI), *LNCS* 2006; 4191: 726-733.

31. Gorbunova V, Sporring J, Lo P, Loeve M, et al. Mass preserving image registration for lung CT. *Medical Image Analysis* 2012; 16: 786-795.

32. Song G, Tustison N, Avants B, Gee JC. Lung CT image registration using diffeomorphic transformation models. In: van Ginneken B, Murphy K, Heimann T, et al, eds. *Medical Image Analysis for the Clinic: A Grand Challenge* 2010; 23-32.

33. Heinrich MP. Deformable lung registration for pulmonary image analysis of MRI and CT scans. PhD thesis, University of Oxford 2013.

34. Cao KL. Mechanical analysis of lung CT images using nonrigid registration. PhD thesis, University of Iowa 2012.

35. Ding K, Cao KL, Fuld MK, Du KF, Christensen GE, et al. Comparison of image registration based measures of regional lung ventilation from dynamic spiral CT with Xe-CT. *Medical Physics* 2012; 39(8): 5084-5098.

36. Murphy K, Ginneken BV, Reinhardt JM, Kabus S, et al. Evaluation of Registration Methods on Thoracic CT: The EMPIRE10 Challenge, *IEEE Transactions on Medical Imaging* 2011; 30: 1901-1920.

37. Allasia G, Cavoretto R, De Rossi A. Local interpolation schemes for landmark-based image registration: a comparison. *Math. Comput. Simulation* 2014; 106: 1-25.

38. Bosica C, Cavoretto R, De Rossi A, Qiao H. On the topology preservation of Gneiting's functions in image registration, *Signal, Image and Video Processing* (2016), to appear; doi: 10.1007/s11760-016-1044-9